\documentstyle[12pt]{article}
\def\beq{\begin{equation}}
\def\eeq{\end{equation}}

\title{ A new scaling property of turbulent flows.}
\author{ R. Benzi$^{1}$, L. Biferale$^1$, S. Ciliberto$^2$,
M.V. Struglia$^1$
and R. Tripiccione$^3$}
\begin{document}
\maketitle
\centerline{$^{1}$  Dipartimento di Fisica, Universit\`{a} ``Tor Vergata''}
\centerline{Via della Ricerca Scientifica 1, I-00133, Rome, Italy.}
\centerline{$^{2}$
 Laboratoire de Physique,  Ecole Normale Sup\'erieure de Lyon }
\centerline{ CNRS URA1325 - 46 All\'ee d'Italie, 69364 Lyon, France}

\centerline{$^{3}$ Istituto Nazionale di Fisica Nucleare, S. Piero a Grado,
Pisa, Italy }

\begin{abstract}
We discuss a possible theoretical interpretation of the self scaling
property of turbulent flows (Extended Self Similarity). Our interpretation
predicts that, even in cases when ESS is not observed, a generalized self
scaling, must be observed. This prediction is checked on a number of
laboratory experiments and direct numerical simulations.
\end{abstract}

\vfill
\newpage

The small scale statistical properties of turbulent flows are usually
described in terms of the probability distribution of the velocity increments
$\delta v(r)\equiv v(x+r)-v(x)$, where $v(x+r)$ and $v(x)$ are velocities along
the $x-$axis at two points separated by a distance $r$.
By assuming locally statistical homogeneity and isotropy and a constant rate
$\epsilon$ of energy transfer from large to small scales, Kolmogorov(1941)
\cite{k41} predicted the existence of an inertial range, i.e. $ \eta\ll r\ll
L$,
$L$ being the integral scale of turbulence and $\eta\equiv\left(\frac{\nu ^3}
{\epsilon}\right)^{1/4}$ the inner or Kolmogorov scale, where the Pdf of
$\delta v(r)$ depends only on $r$ and $\epsilon$.
It then follows that:
\beq
\langle \delta v(r) ^p\rangle \sim \epsilon^{p/3} r^{p/3}
\label{kolmo}
\eeq
Although the basic assumptions behind K41 are usually considered to be
correct, there is rather large evidence that K41 prediction (\ref{kolmo})
is violated in fully developed turbulence, namely one finds in the inertial
range:
\beq
\langle \delta v(r) ^p \rangle \sim r^{\zeta (p)}
\label{anomal}
\eeq
where $\zeta (p)$ is a non linear convex function of $p$ and $\zeta (3)=1$
\cite{urielbook}.
Scaling (\ref{anomal}) is referred to as anomalous scaling because it cannot be
deduced by simple dimensional considerations.

Recently it has been pointed out that scaling (\ref{anomal}) can be generalized
in the following way:
\beq
\langle \delta v(r) ^p \rangle \sim \langle \delta v(r) ^3\rangle^{\zeta ^*(p)}
\label{esscal}
\eeq
where $\zeta ^*(p)\approx \zeta(p)$ \cite{ess}.
Scaling (\ref{esscal}) has been observed
both at low and moderate Reynolds number and for a wider range of scales $r$
with respect to scaling (\ref{anomal})\cite{ess_rev}.
Because of these properties, the self scaling property (\ref{esscal}) of the
velocity field has been named Extended Self Similarity (ESS).
The aim of this Rapid Comm. is to propose an interpretation of ESS.
Moreover, our interpretation predicts a generalized form of ESS which should
hold also for non isotropic and non homogeneous turbulence and for any scale
$r$.
These predictions are supported by experimental and numerical results.

Our starting point is the multifractal interpretation of anomalous scaling
(\ref{anomal}), namely:
\beq
\langle \delta v(r) ^p \rangle \sim \int d\mu (h) r^{hp} r^{3-D(h)}
\label{multifrac}
\eeq
\beq
\zeta (p) = \inf _{h} [ hp + 3-D(h)]
\label{multifrexp}
\eeq
where $D(h)$ is assumed to be the fractal dimension of the set of points where
$\delta v(r) \sim r^h$.\\
Many phenomenological multifractals models for $D(h)$ have been proposed.
Among them, we shall consider those models which are consistent with a
infinitively divisible distribution of random multiplier \cite{idd}
\cite{kolmo_ref}.
For all these models
$D(h)$ can be written as:
\beq
D(h) = 3 - d_0 f\left(\frac{h-h_0}{d_0}\right)
\label{ddh}
\eeq
Different models give different shape of the function $f(x)$ and suggest
different physical interpretation of $h_0$ and $d_0$.

By using (\ref{ddh}) into (\ref{multifrexp}) we obtain:
\beq
\zeta(p) = h_0 p + d_0 H(p)
\label{zetap}
\eeq
where
\beq
 H(p) = \inf _x \left[px + f(x)\right]
\label{Hp}
\eeq
In order to clarify the following discussion, let us consider the She-Leveque
model which is in remarkable good agreement with existing experimental and
numerical data \cite{sl}.
In this model $h_0$ characterizes the most singular behaviour of the
velocity field and $D_0\equiv 3-d_0$ the corresponding fractal dimension.

At low Reynolds number or, equivalently, at small scales $r$, the effect of
viscosity $\nu$ may become relevant. In simple phenomenological models, the
effect of viscosity is usually represented as a cutoff in the energy transfer.
Here we consider an alternative point of view: the energy transfer, as well as
its fluctuations responsible for intermittency effect, continues to hold and,
because of viscosity, the probability
distribution of the velocity increments acquires a dependence
on the ratio $\left( r/\eta\right)$. If this is the case, both $h_0$ and $d_0$
may acquire a (smooth) dependence on $r$.
Indeed, we expect that the role of the viscosity should increase the value of
$h_0$ (i.e. reduce the strength of maximum singularity)
and to reduce the number
of structures where $\delta v(r)\sim r^{h_0(r)}$.
Thus, the probability $P$ to observe a local scaling
$\delta v(r)\sim r^{h_0(r)}$
should decrease. Because $P\sim r^{3-D_0}\sim r^{d_0}$, we deduce that $d_0$
should be an increasing function of $r$.

If our picture is qualitatively correct, ESS simply states that $\frac{h_0(r)}{
d_0(r)}=const$, i.e. the dependence on $r$ of $h_0$ and $d_0$ is the same.
Indeed, one has:
$$
\frac{\zeta(p)}{\zeta(q)} =
\frac{\frac{h_0}{d_0} p + H(p)}{\frac{h_0}{d_0} q + H(q)}
$$
which does not depend on $r$.
Let us remark that a smooth dependence on $r$ of $h_0$ and $d_0$ does not spoil
the saddle point integration (\ref{multifrac}) on $d\mu (h)$.\\ Also,
let us note that this interpretation of ESS allows us to generate
synthetic turbulence signal, by random multiplicative process, which
shows ESS.
Eventually at very small scale the effect of viscosity is strong enough to
destroy ESS. In homogeneous and isotropic turbulence, ESS is indeed broken
at small scales of order of few ($5\div 6$) Kolmogorov length \cite{ess_rev}.

Our interpretation of ESS is based upon the assumptions that the statistical
properties of turbulence at low $Re$ or at small scales are controlled by
(\ref{multifrac}) with $h_0$ and $d_0$ smooth functions of $r$
and $h_0/d_0=const$.
This implies a (delicate) balance between the scaling of the most
singular
structures in a turbulent flow and the number of these structures. This balance
can be broken in different ways. For instance, near boundary layers or in
strong shear flow conditions, energy production and momentum transfer can
significantly change the slope and the number of the most singular structures.
In these cases ESS should not be observed \cite{ess2} \cite{sreenivasan}.
However, even in cases where ESS is not observed, our theoretical
interpretation
could still be  valid and we think it is very important to check
any possible prediction.

To this aim, let us consider the following dimensionless quantity:
\beq
G_p(r) = \frac{\langle \delta v(r) ^p \rangle}{\langle \delta v(r) ^3\rangle^
{p/3}}
\label{adimen}
\eeq
According to (\ref{zetap}) and (\ref{Hp}) we obtain:
\beq
G_p(r) = r^{d_0[H(p)-\frac{p}{3}H(3)]}
\label{Spscal}
\eeq
Our theoretical interpretation of ESS suggest that $G_p(r)$ should always
satisfy the self scaling properties:
\beq
G_p(r)\sim G_q(r)^{\rho (p,q)}
\label{sess}
\eeq
regardless of any boundary layer, shear flow or viscosity which can spoil ESS.
In (\ref{sess}) $\rho (p,q) = \left( H(p)-\frac{p}{3}H(3) \right) /\left( H(q)
-\frac{q}{3}H(3)\right)$ and does not depend on $d_0$.\\
We have checked (\ref{sess}) in a variety of turbulent flows. We have found
that (\ref{sess}) is always satisfied within the accuracy of statistical
errors.
In Fig.1 we plot $G_6(r)$ against $G_5(r)$, in a log-log scale,
for few cases three of which do not
show ESS.

In all cases, we have found that (\ref{sess})
is satisfied down to the smallest
scale available in our laboratory experiments or numerical simulations.
Also, we have found that (\ref{sess}) holds also in the limit where
$\delta v(r) \sim r$. This means that, in terms of the self-scaling properties
of $G_p(r)$, no evidence of a viscous cutoff has been observed.

The validity of (\ref{sess})  (which we refer to as Generalized ESS) may have
important theoretical consequence. Indeed, it has been observed in
\cite{ess_rev} \cite{journaldephysique1},
that the following ESS form of the Kolmogorof
Refined Similarity Hypothesis is always
satisfied in turbulent flows:
\beq
\langle \delta v(r) ^{p}\rangle\sim \langle \epsilon (r)^{p/3}\rangle
\langle \delta v(r) ^3\rangle ^{p/3}
\label{RSH}
\eeq
where $\epsilon (r)$ is defined as the local energy dissipation averaged on
a box of side $r$. Because of (\ref{RSH}), equation (\ref{sess}) tells us that
$\epsilon (r)$ displays ESS on all scales regardless the effect of boundary
layers and shear flows.
This gives strong constraints on how a turbulent flow can dissipate energy on
small scales.
In particular, viscous effects do not change the anomalous scaling
in $\epsilon (r)$ in any appreciable way, at variance with existing theoretical
and phenomenological model of turbulence \cite{urielbook}
\cite{fv}.

A more systematic presentation of our results, including a simple model to
generate statistical signal in agreement with equations (\ref{sess})
and (\ref{RSH}), is under preparation.

Discussion with G. Stolovitzky and S. Fauve
are kindly acknowledged. L.B. has been
partially supported by the CEE contract ERBCHBICT941034.
This work was partially supported by the CEE contract CT93-EV5V-0259.

\vfill
\newpage
{\bf Figure Captions.}
\begin{description}
\item[Figure 1.]
The figure shows $\log S_6(r)$ plotted against $\log S_5(r)$ for $4$ different
experimental and numerical cases. (0) refers to the wake of a cylinder of $10$
cm diameter taken at $60$ cm downstream. In this case no ESS is observed (see
\cite{ess2} for further details).
Diamonds refers to the hot wire measurement taken at
$z=7$ mm of a boundary layer (courtesy of G.R. Chavarria). Also in this case
ESS is not observed. Squares refers to a Direct Numerical Simulation
of turbulent convection at $Ra \approx 10^7$ \cite{bolgiano}.
Crosses refers to direct numerical simulation of
a Kolmogorof flow \cite{bs} at $Re_{\lambda} \approx 40$.

\end{description}
\vfill
\end{document}